\documentclass[acs,superscriptaddress,floatfix,10pt]{revtex4-1}
\usepackage{graphicx}
\usepackage{dcolumn}
\usepackage{latexsym}
\usepackage{cleveref}
\usepackage{amsmath}
\usepackage{changes}
\begin{document}
\title{Fidelity of bacterial translation initiation: a stochastic kinetic model}
\author{Dipanwita Ghanti} 
\affiliation{Department of Physics, Indian Institute of Technology, Kanpur 208016, India} 
\author{Kelvin Caban} 
\affiliation{Department of Chemistry, Columbia University, New York, NY 10027, USA}
\author{Joachim Frank} 
\affiliation{ Department of Biochemistry and Molecular Biophysics
and Department of Biological Sciences, New York, NY 10032, USA}
\author{Ruben L. Gonzalez, Jr.}
\affiliation{Department of Chemistry, Columbia University, New York, NY 10027, USA}
\author{Debashish Chowdhury}
\affiliation{Department of Physics, Indian Institute of Technology, Kanpur 208016, India}

\begin{abstract}
During the initiation stage of protein synthesis, a ribosomal initiation complex (IC) is assembled on a messenger RNA (mRNA) template. In bacteria, the speed and accuracy of this assembly process are regulated by the complementary activities of three essential initiation factors (IFs). Selection of an authentic N-formylmethionyl-transfer RNA (fMet-tRNA\textsuperscript{fMet}) and the canonical, triplet-nucleotide mRNA start codon are crucial events during assembly of a canonical, ribosomal 70S IC. Mis-initiation due to the aberrant selection of an elongator tRNA or a non-canonical start codon are rare events that result in the assembly of a pseudo 70S IC or a non-canonical 70S IC, respectively. Here, we have developed a theoretical model for the stochastic kinetics of canonical-, pseudo-, and non-canonical 70S IC assembly that includes all of the major steps of the IC assembly process that have been observed and characterized in ensemble kinetic-, single-molecule kinetic-, and structural studies of the fidelity of translation initiation. Specifically, we use the rates of the individual steps in the IC assembly process and the formalism of first-passage times to derive exact analytical expressions for the probability distributions for the assembly of  canonical-, pseudo- and non-canonical 70S ICs. In order to illustrate the power of this analytical approach, we compare the theoretically predicted first-passage time distributions with the corresponding computer simulation data. We also compare the mean times required for completion of these assemblies with experimental estimates. 
In addition to generating new, testable hypotheses, our theoretical model can also be easily extended as new experimental 70S IC assembly data become available, thereby providing a versatile tool for interpreting these data and developing advanced models of the mechanism and regulation of translation initiation.
\end{abstract}
\maketitle
\section{Introduction}

Over the past fifteen years, enormous progress has been made towards understanding the mechanisms of operation of 
a large number molecular machines \cite{frank11}. 
Most recently, ground-breaking advances in the development of synthetic 
 molecular machines have triggered the a ``nano-scale industiral revolution" \cite{nobel}. 
 Despite this progress, the fundamental differences between  molecular machines 
and their macroscopic counterparts  makes studies of molecular machines so important and interesting. 
For example, assembly of a molecular machine typically involves biasing of a series of stochastic, reversible interactions between multiple molecular components, a process that is very different from the deterministic procedure with which a macroscopic machine such as a car engine is assembled.
 Timely and successful operation of such a molecular machine requires rapid and accurate assembly.
 Delayed assembly can stall a biological function, while defective assembly can cause malfunction; both of these can have disastrous effects and can even endanger the survival of the cell. 
Consequently, understanding how molecular machines are rapidly and accurately assembled is  one of the
most important, and yet most challenging, frontiers in research on molecular machine.
 The theoretical approach that we develop here for modeling the stochastic assembly kinetics of a molecular machine
  is  complementary to the experimental studies of this process.

Proteins are the workhorses of a living cell \cite{tanford01} and amino acids are the monomeric subunits of 
these hetero-polymeric molecules. A protein is synthesized by the addition of successive amino acids, as 
directed by the corresponding messenger RNA (mRNA) template. Each mRNA is also a hetero-polymer 
consisting of a sequence of nucleotides that are its monomeric subunits. The successive triplets of nucleotides, 
called codons, within the open reading frame of the mRNA determines the corresponding sequence of amino acids
to be added to the nascent protein.
This mRNA-directed synthesis of a protein, called translation (of the genetic message supplied by an mRNA), is carried out by a ribosome \cite{frank11,spirin00,rodnina12}. Each ribosome is a large, two-subunit, macromolecular machine that
is composed of a large number of distinct ribosomal proteins and ribosomal RNA (rRNA) molecules. Amino acids are delivered
to the ribosome by transfer RNA (tRNA) adapter molecules, which contain a triplet-nucleotide anti-codon that pairs with its cognate mRNA codon. In bacteria, the process of translation begins with the assembly of a ribosomal 70S  initiation complex (IC) at the ribosome binding site of an mRNA. One of the steps in 70S IC assembly process involves the hydrolysis of a high-energy phosphate bond in guanosine triphosphate (GTP). Consequently, the IC assembly process
consumes energy and the assembled 70S IC need not be in thermodynamic equilibrium.  
Such a process is often referred to as a `self-organization' process, to distinguish it from a `self-assembly' process
that, by definition, leads to the stable equilibrium structure of a multicomponent supra molecular aggregate \cite{misteli01}.

Just like most of the kinetic processes in molecular biology, the assembly of an IC is intrinsically stochastic 
\cite{bressloff14}. Consequently , the time taken for assembly to complete as well as the fidelity of assembly vary randomly from
one 70S IC to other. In this article, we develop a theoretical model for the stochastic kinetics of 70S IC assembly that includes all of the major pathways  (\textit{i.e.}, the canonical, pseudo, and non-canonical 70S IC pathways) and all of the key steps along each pathway. We formulate the kinetics of the model in terms of master equations for the appropriately  defined probabilities. We analyze these equations using the techniques of stochastic processes and non-equilibrium statistical mechanics. For a stochastic process, the time taken by the system to attain a specific state {\it for the first time}, starting from another specific given initial state, is called the corresponding, first-passage time \cite{redner01}. Distribution of first-passage times are central features of stochastic process in molecular biology \cite{bressloff14,metzler14,biswas16}. The complete assembly of a 70S IC, starting from well 
defined-initial configurations of its components, can be formulated as a first-passage problem. We calculate exact analytical expressions for 70S IC assembly along the canonical, pseudo, and non-canonical pathways such that the
the corresponding mean first-passage times reflect the average times required for 70S IC assembly along the respective pathways.

\section{Mechanism and Regulation of translation initiation}

The bacterial 70S ribosome  consists of two subunits, a small, or 30S, subunit (SSU) and a large, or 50S, subunit (LSU).
Guided by protein translation factors, GTP, and aminoacyl-tRNA (aa-tRNA) substrates, the ribosome translates an
mRNA into protein in three distinct stages termed: initiation, elongation and termination. One of three canonical
`start' codons (most frequently AUG, but also GUG and UUG) on the mRNA serves as the genetically encoded signal for the 
initiation of translation, while three `stop' codons (UGA,UAG, and UAA) downstream along the mRNA signals termination of the process. Initiation, which is the rate limiting stage of the overall process of protein synthesis , 
 is a highly regulated, multi-step process that results in the assembly of a mature, elongation 
competent, 70S IC. The canonical 70S IC contains an initiator, N-formyl-methionyl-tRNA (fMet-tRNA\textsuperscript{fMet}) that is base-paired to the start codon of the mRNA at the ribosomal peptidyl-tRNA binding (P) site. Following assembly of the 70S IC, the second codon on the mRNA is positioned in the ribosomal aminoacyl-tRNA binding (A) site and the 70S IC is primed to enter into the elongation stage.
The elongation stage is a cyclical process that involves three sequential, repetitive steps: 1) decoding of the A-site 
codon by a cognate aa-tRNA substrate, 2) peptide bond formation between the amino acid moeity of the aa-tRNA in the A site and
the fMet or peptidyl moiety on the fMet-tRNAfMet or peptidyl-tRNA, respectively, in the P site and 3) `translocation', a three- nucleotide movement of the ribosomal 70S elongation complex along the mRNA, which brings the next
codon into the A site. The termination stage begins when a stop codon is translocated into the A site and is
 decoded by a release factor that binds to the A site and catalyzes the release of fully synthesized protein.

The assembly of the 70S IC is kinetically regulated by three essential initiation factors (IF1, the translational guanosine triphosphatase IF2, and IF3) \cite{gualerzi15}, and occurs in two distinct, major phases. In the first major phase, a 30S IC composed of the 30S subunit, the three IFs, and the initiator fMet-tRNA\textsuperscript{fMet}, is assembled at the ribosome binding site of the mRNA. In the second major phase, the 50S subunit docks onto the 30S IC to form a 70S IC* (in which the * denotes a 70S IC that is not yet ready to enter the elongation stage of protein synthesis).
. Following a series of subsequent sub-steps, including the hydrolysis of GTP by IF2, the dissociation of the IFs, clockwise rotation of the 30S subunit relative to the 50S subunit (as viewed from the 30S solvent face),
 and the placement of the fMet-tRNA\textsuperscript{fMet} into the peptidyl transferase center, the 70S IC* matures into an elongation-competent form. The arrival times of the IFs, extracted from ensemble rapid kinetic data, have revealed that, during the assembly of the 30S IC, IF3 and IF2 are the first to bind the 30S subunit, followed by IF1, which increases the affinity and amplifies the activities of the other two factors \cite{milon12,antoun06,antounEMBO06}. Consequently, the 30S IC-bound IFs act synergistically to regulate the kinetics of tRNA binding such that fMet-tRNA\textsuperscript{fMet} is selectively recruited into the start codon-programmed P site of the 30S IC. In addition, the 30S IC-bound IFs synergistically tune the kinetics of 50S subunit docking such that the 50S subunit preferentially docks onto a 30S IC that has a P site-bound fMet-tRNA\textsuperscript{fMet} that is base- paired to a canonical start codon. IF3 functions by uniformly increasing the dissociation of tRNAs from the 30S P site \cite{antoun06} and, together with IF1, by antagonizing the docking of the 50S subunit to 30S ICs that either lack an authentic fMet-tRNA\textsuperscript{fMet}, or that have an authentic fMet-tRNA\textsuperscript{fMet} that is base-paired to a non-canonical start codon \cite{milon08}. IF2 functions by selectively accelerating the binding of fMet-tRNA\textsuperscript{fMet} into the 30S P site and, subsequently, by selectively accelerating 50S subunit docking to 30S ICs that contain and fMet-tRNA\textsuperscript{fMet} that is base-paired to a start codon \cite{antoun06}. 
 IF2 and fMet-tRNA\textsuperscript{fMet} form a sub-complex on the inter-subunit face of the 30S IC \cite{simonetti08,hussain16}
 and this interaction enables the GTP-form of IF2 to acquire a conformation that is active for rapid 50S subunit docking \cite{zorzet10,pavlov11,caban17}. 

Like any other process of assembling a multi-component macromolecular complex in a living cell, that of assembling the 70S IC also suffers from the occasional error. Two aberrant 70S ICs have also been observed \textit{in vivo} and \textit{in vitro},
a pseudo 70S IC \cite{samhita13,guillon92,zorzet10,pavlov11,antoun06} and a non-canonical 70S IC \cite{sussman96,sacerdot96,haggerty97,olsson96,qin12,grigoriadou07,milon08} each of which is caused by a distinct mis-initiation event.
 The assembly of 
a pseudo 70S IC is caused by an error in the selection of the authentic fMet-tRNA\textsuperscript{fMet} (e.g., the 70S IC carries either a non-formylated Met-tRNA\textsuperscript{fMet} or an elongator aa-tRNA). Conversely, the assembly of a non-canonical 70S IC results from an error in the selection of the mRNA start codon (e.g., selection of an AUU near-start codon). 

Initiation \textit{via} the non-canonical 70S IC pathway is inefficient \textit{in vivo} supporting translation at a level that is $1-3\%$ or less of the canonical pathway\cite{sussman96}. However, inactivating mutations in the \textit{infC} gene encoding IF3\cite{sussman96,sacerdot96,haggerty97}, a reduction in the expression of IF3 \cite{olsson96}, and mutations in the 16S ribosomal RNA (rRNA) component of the 30S subunit \cite{qin12}, have all been shown to increase initiation \textit{via} the non-canonical 70S IC pathway. Initiation \textit{via} the pseudo 70S IC pathway with an elongator aa-tRNA has only been observed in cells when tRNA\textsuperscript{fMet} levels were reduced by $ \sim 75 \% $ \textit{via} deletion of the \textit{met}ZWV locus containing three of the four initiator tRNA genes\cite{samhita13}. 
Initiation \textit{via} the pseudo 70S IC pathway with a non-formylated Met-tRNA\textsuperscript{fMet} is thought to occur in cells containing an inactivating mutation in the \textit{fmt} gene encoding the formyl-methionine transformylase enzyme \cite{guillon92}. Highlighting the importance of initiation \textit{via} the canonical pathway, \textit{fmt} null cells exhibit a severe growth defect. Extragenic suppressors of the \textit{fmt} mutation have been identified in the \textit{infB} gene encoding IF2 \cite{zorzet10}. These suppressor mutations result in an error-prone variant of IF2 that has been shown to initiate with non-formylated Met-tRNA\textsuperscript{fMet}, with elongator aa-tRNAs, and with no tRNA at all \textit{in vitro} and, presumably, \textit{in vivo} \cite{zorzet10,pavlov11}. 
 
  Almost all of the kinetic models of translation that have been reported so far \cite{basu07,garai09,sharma11,dutta17,bar07,zouridis07}, have mainly focused on the details of the elongation stage of protein synthesis and have represented the initiation stage of protein synthesis using a single effective rate constant. Here, we have developed a kinetic model that focuses exclusively on the multi-step assembly of the 
canonical, pseudo, and non-canonical 70S ICs during the initiation stage of protein synthesis.
The process of 70S IC assembly can be modeled at different levels of spatial and temporal resolution depending on the specific phenomena that the model is intended to describe. The model developed here is not intended to include assembly of the 30S and 50S 
 subunits from their respective constituent ribosomal proteins and rRNA molecules. It is also not aimed to describe the
 process through which the 30S subunit or 30S IC searches for the ribosome binding site on the mRNA to be translated.
   Instead the initial state in our model is one in which the 30S subunit is already
bound to all the IFs and the model focuses on the subsequent IF- regulated selection 
of the start codon and fMet-tRNA\textsuperscript{fMet} and the ensuing IF-regulated docking of the 50S subunit.

\section{Model}

\begin{figure}[htb]
\center
\includegraphics[angle=0,width=1.0\textwidth]{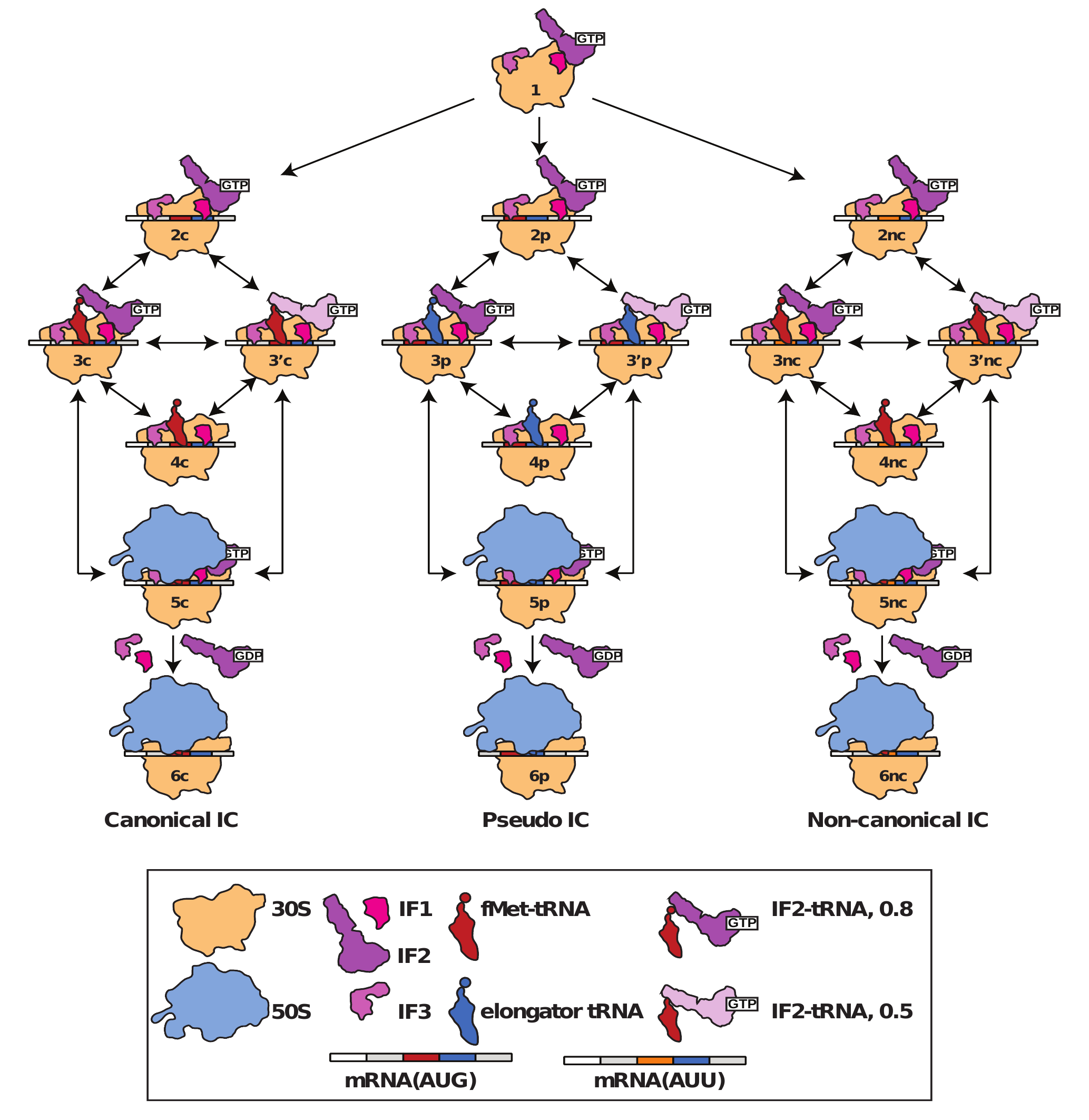}
\caption{A schematic representation of the three alternative, multi-step kinetic pathways for the assembly of 70S IC.
During canonical 70S IC assembly (pathway along leftmost scheme), an fMet-tRNA\textsuperscript{fMet} forms base-pairing interactions with a canonical AUG start codon.
During pseudo 70S IC assembly (pathway along center scheme), a non-formylated Met-tRNA\textsuperscript{fMet} or an elongator aa-tRNA forms base-pairing interactions with a cognate sense codon as shown in the figure.
During non-canonical 70S IC assembly (pathway along rightmost scheme), an fMet-tRNA\textsuperscript{fMet} attempts to form base-pairing interactions with a non-canonical AUU near-start codon.} 
\label{fig_model}
\end{figure}

The sequence of association and dissociation events along the three possible 70S IC pathways is shown in Figure\ref{fig_model}. In our model, all three pathways begin at state `1' with a 30S subunit \textit{i.e.} pre-bound with the three IFs. In state `2c' of the canonical pathway, the 30S IC binds to an mRNA and positions the AUG start codon in the P site and `c' indicates the pathway is canonical.
 The transition $ 1\rightarrow 2c $ takes place at a rate of  $ k_{12c} $. In state `3c', fMet-tRNA\textsuperscript{fMet} binds to the P site of the 30S IC such that  its anti-codon base-pairs with the start codon and its aminoacyl acceptor stem interacts with IF2, resulting in the formation of an IF2-tRNA sub complex on the 30S IC. The transition $ 2c\rightarrow 3c $ takes place at a rate of $ k_{2c3c} $. Previous single-molecule fluorescence resonance energy transfer (smFRET) studies have shown that the IF2-tRNA sub-complex on the 30S IC complex
reversibly fluctuates between two
  conformational states with FRET efficiencies ($ E_{FRETS} $) of $\sim $0.8 and $\sim$0.5; in our model, these states are labeled by $ 3c $ and $ 3'c $,  respectively \cite{wang15}. Dissociation and re-association of IF2 is captured by the transition  $ 3c\rightleftharpoons 4c $ and  $ 3'c\rightleftharpoons 4c $. Docking of the 50S subunit to a 30S IC in either state $ 3c $ or $ 3'c $ represents the next transition, $ 3c\rightarrow 5c $ or $ 3'c\rightarrow 5c $. Subsequent GTP hydrolysis and release of all the IFs, which completes formation of the canonical 70S IC, is indicated by the transition $ 5c\rightarrow 6c $.

In the pseudo 70S IC pathway, the process indicated by the transition $ 1\rightarrow 2p $ is identical to that in the transition $ 1\rightarrow 2c $ along the canonical pathway except that an elongator aa-tRNA is positioned in the 30S IC P site. The `p' indicates the pathway is pseudo. The next transitions $ 2p \rightarrow3p $ and $ 2p \rightarrow3'p $, are similar to the transitions $ 2c\rightarrow 3c $ and $ 2c\rightarrow 3' c$, respectively, along the canonical pathway except that the elongator aa-tRNA now attempts to base-pairs to a cognate codon proximal to the AUG start codon. 
The two dynamic conformational states $ 3p $ and $ 3'p $ are the analogs of the states $3c$ and $3'c$ on the canonical pathway. During fluctuations between the states $ 3p $ and $ 3'p$, IF2 can dissociate from the 30S IC; this process is represented by the transition to the state $ 4p $, which results in a premature abortion of the 70S IC assembly process. Alternatively, the 50S subunit can successfully dock onto the 30S IC in state $3p$ or $3'p$, indicated by the transition $ 3p\rightarrow 5p$ or $ 3'p\rightarrow 5p$. Following the hydrolysis of GTP and release of all the IFs, represented by the transition $5p\rightarrow 6p$, assembly of the pseudo 70S IC is completed.

In the non-canonical 70S IC pathway, the fMet-tRNA\textsuperscript{fMet} attempts to base-pair with a non-canonical AUU near-start codon, rather than an AUG start codon, during the transition $ 2nc\rightarrow 3nc $ and $2nc\rightarrow 3'nc $,  `nc' indicates the pathway is non-canonical. The remaining states and transitions are analogous to those defined in the canonical and pseudo 70S IC pathways. 

In Figure\ref{fig_model} transitions connecting states directly using a two-headed arrow in mutually opposite directions represent reversible transitions whereas irreversible transitions are denoted by connecting a pair of states by a single-headed arrow. The symbol $ k_{ij} $, is the probability of a transition from state $i$ to state $j$ per unit time; we'll refer to these, loosely speaking, as the `rates' of the corresponding transitions. $ P_{i}(t) $ denotes the probability that, at a given instant of time $ t $, the system is in the state $ i $. Time evolution of these probabilities are governed by the corresponding master equations (given in the Appendix). For the analytical calculation of the distributions of the first-passage times (FPTs), we solve these master equations for the initial condition 

\begin{eqnarray}
P_{1}(0)=1 \ & and & \  P_{i}(0)=0 , i \neq 1. \
\label{eq_initial_cond}
\end{eqnarray}
Because of the normalization condition, $ \sum_{i}P_{i}(t)=1 $, not all of the $P_i(t)$ are independent of each other.
Formally, all the master equations can be expressed in a compact form using the matrix notation 
\begin{equation}
\frac{d\widehat{P}(t)}{dt}=A\widehat{P}(t),
 \label{eq_master}
\end{equation}

where $ \widehat{P}(t) $ is a column matrix consisting of $16$ elements, while the elements of the $16 \times 16$ matrix $A$ are the various rate constants involved in the master equations. Formal solution of these master equations is given by $\widehat{P}(t)=e^{A.t}\widehat{P}(0)$,  where $\widehat{P}(0)$ is the column matrix that denotes the initial condition. Suppose $f_{c}(t) \Delta t$ is the probability for the completion of a canonical 70S IC (\textit{i.e.}, passage to the 
state $5$ for the first time) between times $t$ and $t + \Delta t$. Similarly, we also define the corresponding probabilities $f_{p}(t) \Delta t$ and $f_{nc}(t) \Delta t$ for the completions of the assemblies of the pseudo 70S IC and non-canonical 70S IC, respectively, between times $t$ and $t+\Delta t$. It is straightforward to show that, for the initial conditions (\ref{eq_initial_cond}), these three distributions 
of the FPTs are given by 
\begin{eqnarray}
f_{c}(t)=k_{5c6c}P_{5c}(t), ~~
f_{p}(t)=k_{5p6p}P_{5p}(t), ~~\text{and }
f_{nc}(t)=k_{5nc6nc}P_{5nc}(t). \nonumber \\
\label{eq_fpt}
\end{eqnarray}

Using the solutions of the master equations, corresponding to the initial conditions (\ref{eq_initial_cond}), 
we get the three FPT distributions from which obtain the corresponding mean FPTs from the defining relations
\begin{eqnarray}
<t_{c}>=\int_{0}^{\infty}tf_{c}(t)dt, ~~
<t_{p}>=\int_{0}^{\infty}tf_{p}(t)dt, ~~\text{and }
<t_{nc}>=\int_{0}^{\infty}tf_{nc}(t)dt. \nonumber\\
\label{eq_tavg}
\end{eqnarray}
The inverse of these mean FPTs can be interpreted as the rates $ k_{c},k_{p} $ and $ k_{nc} $ for 
assembling the respective 70S ICs.
The randomness parameter, which quantifies the relative strength of the fluctuation and mean of the FPT is given by 
\begin{eqnarray}
r_{c}=\frac{<t^{2}_{c}>-<t_{c}>^{2}}{<t_{c}>^{2}}, ~~
r_{p}=\frac{<t^{2}_{p}>-<t_{p}>^{2}}{<t_{p}>^{2}}, ~~\text{and }
r_{nc}=\frac{<t^{2}_{nc}>-<t_{nc}>^{2}}{<t_{nc}>^{2}}.\nonumber\\
\label{eq_randomness}
\end{eqnarray}
where 
\begin{eqnarray}
<t^{2}_{c}>=\int_{0}^{\infty}t^{2}f_{c}(t)dt, ~~
<t^{2}_{p}>=\int_{0}^{\infty}t^{2}f_{p}(t)dt, ~~ \text{and }
<t^{2}_{nc}>=\int_{0}^{\infty}t^{2}f_{nc}(t)dt. \nonumber\\
\end{eqnarray} 

For formation of a 70S IC that is completed at any arbitrary instant of time $t$, the probability that it is canonical, pseudo or non-canonical is given by the probabilities ${\cal P}_{c}$, ${\cal P}_{p}$ and ${\cal P}_{nc}$, respectively, where
\begin{eqnarray}
{\cal P}_{c}&=&\frac{k_{5c6c}P_{5c}(t)}{k_{5c6c}P_{5c}(t)++k_{5p6p}P_{5p}(t)+k_{5nc6nc}P_{5nc}(t)} \nonumber \\
{\cal P}_{p}&=&\frac{k_{5p6p}P_{5p}(t)}{k_{5c6c}P_{5c}(t)++k_{5p6p}P_{5p}(t)+k_{5nc6nc}P_{5nc}(t)} \nonumber \\
{\cal P}_{nc}&=&\frac{k_{5nc6nc}P_{5nc}(t)}{k_{5c6c}P_{5c}(t)++k_{5p6p}P_{5p}(t)+k_{5nc6nc}P_{5nc}(t)} \nonumber \\
\label{eq_fraction}
\end{eqnarray}

\section{Parameters}

\begin{table}[h]
\caption{Values of different parameters\label{table_parameter}}
\begin{center}

\begin{tabular}{|c|c|c|c|}
\hline 
\multicolumn{1}{|p{2cm}}{\centering for canonical 70S IC }
& \multicolumn{1}{|p{2cm}}{\centering for pseudo 70S IC}
& \multicolumn{1}{|p{2cm}|}{\centering for noncanonical 70S IC}\\
\hline
$k_{12c}=$ 122$ \mu M^{-1}s^{-1}$\cite{studer06} & $k_{12p}=$122$\mu M^{-1}s^{-1}$\cite{antoun06} &$k_{12nc}=$ 93$ \mu M^{-1}s^{-1}$\cite{studer06}\\
\hline
$k_{2c3c}=$ 12.5$\mu M^{-1}s^{-1}$\cite{antoun06}& $k_{2p3p}=$ 0.22$\mu M^{-1}s^{-1}$\cite{antoun06} &$k_{2nc3nc}=$ 12.5$\mu M^{-1}s^{-1}$\cite{antoun06}\\
\hline 
$k_{2c3'c}=$ 12.5$\mu M^{-1}s^{-1}$\cite{antoun06} & $k_{2p3'p}=$ 0.22$\mu M^{-1}s^{-1}$\cite{antoun06} &$k_{2nc3'nc}=$ 12.5$\mu M^{-1}s^{-1}$\cite{antoun06}\\
\hline
$k_{3c2c}=$ 0.034$s^{-1}$\cite{antoun06} & $k_{3p2p}=$ 0.42$ s^{-1} $ \cite{antoun06}&$k_{3nc2nc}=$ 0.034$s^{-1}$\cite{antoun06}\\
\hline 
$k_{3'c2c}=$ 0.034$s^{-1}$\cite{antoun06} & $k_{3'p2p}=$  0.42$ s^{-1} $ \cite{antoun06}&$k_{3'nc2nc}=$ 0.034$ s^{-1} $\cite{antoun06}\\
\hline 
$k_{3c3'c}=$ 1.3$s^{-1}$\cite{wang15} & $k_{3p3'p}=$ 0.1$ s^{-1} $ \footnote{Not available}&$k_{3nc3'nc}=$ 0.1$ s^{-1} $ \textsuperscript{a}\\
\hline 
$k_{3'c3c}=$ 0.96$s^{-1}$\cite{wang15} & $k_{3'p3p}=$ 0.1$ s^{-1} $ \textsuperscript{a} &$k_{3'nc3nc}=$ 0.1$s^{-1}$ \textsuperscript{a}\\
\hline
$k_{3c4c}=$ 0.013$s^{-1}$ \cite{wang15}& $k_{3p4p}=$ 0.1$s^{-1}$ \textsuperscript{a} &$k_{3nc4nc}=$ 0.65$s^{-1}$ \cite{wang15}\\
\hline 
$k_{3'c4c}=$ 0.013$ s^{-1}$ \cite{wang15} & $k_{3'p4p}=$ 0.1$s^{-1}$ \textsuperscript{a} & $k_{3'nc4nc}=$ 0.65$s^{-1}$ \cite{wang15}\\
\hline
$k_{4c3c}=$ 13$\mu M^{-1}s^{-1}$\cite{wang15} &$k_{4p3p}=$ 0.1$\mu M^{-1}s^{-1}$ \textsuperscript{a} &$k_{4nc3nc}=$ 4.2$\mu M^{-1}s^{-1}$ \cite{wang15}\\
\hline
$k_{4c3'c}=$13$\mu M^{-1}s^{-1}$ \cite{wang15} &$k_{4p3'p}=$ 0.1$\mu M^{-1}s^{-1}$ \textsuperscript{a} &$k_{4nc3'nc}=$ 4.2$\mu M^{-1}s^{-1}$ \cite{wang15}\\
\hline
$k_{3c5c}=$ 2.9$s^{-1}$ \cite{antounEMBO06} & $k_{3p5p}=$ 0.0071$s^{-1}$\cite{antoun06} &$k_{3nc5nc}=$ 0.06$s^{-1}$ \cite{wang15}\\
\hline 
$k_{3'c5c}=$ 2.9$ s^{-1}$ \cite{antounEMBO06}  & $k_{3'p5p}=$ 0.0071$s^{-1}$ \cite{antoun06}& $k_{3'nc5nc}=$ 0.06$s^{-1}$\cite{wang15} \\
\hline
$k_{5c3c}=$ 1.12$s^{-1}$\cite{macdougall15} &$k_{5p3p}=$ 3$s^{-1}$ \textsuperscript{a} &$k_{5nc3nc}=$ 3$s^{-1}$ \textsuperscript{a}\\
\hline
$k_{5c3'c}=$1.12$s^{-1}$ \cite{macdougall15}  &$k_{5p3'p}=$ 3$s^{-1}$ \textsuperscript{a} &$k_{5nc3'nc}=$ 3$s^{-1}$ \textsuperscript{a}\\
\hline
$k_{5c6c}=$ 1.76$s^{-1}$ \cite{goyal15}& $k_{5p6p}=$1.76$s^{-1}$\cite{goyal15} \footnote{This value is taken from Ref.\cite{goyal15}, assuming this value is identical to $k_{5c6c}$}&$k_{5nc6nc}=$ 1.76$s^{-1}$ \cite{goyal15} \textsuperscript{b}\\
\hline
\end{tabular}
\end{center}
\end{table}

\begin{table}[h]
\caption{Values of different parameters used for calculation \label{table_used}}
\begin{center}
\begin{tabular}{|c|c|c|c|}
\hline 
\multicolumn{1}{|p{2cm}}{\centering for canonical 70S IC }
& \multicolumn{1}{|p{2cm}}{\centering for pseudo 70S IC}
& \multicolumn{1}{|p{2cm}|}{\centering for noncanonical 70S IC}\\
\hline
$k_{12c}=$ 18.3$ s^{-1}$\cite{studer06} & $k_{12p}=$18.3$s^{-1}$\cite{antoun06} &$k_{12nc}=$ 13.95$s^{-1}$\cite{studer06}\\
\hline
$k_{2c3c}=$ 1.875$s^{-1}$\cite{antoun06}& $k_{2p3p}=$ 0.033$s^{-1}$\cite{antoun06} &$k_{2nc3nc}=$ 1.875$s^{-1}$\cite{antoun06}\\
\hline 
$k_{2c3'c}=$ 1.875$s^{-1}$\cite{antoun06} & $k_{2p3'p}=$ 0.033$s^{-1}$\cite{antoun06} &$k_{2nc3'nc}=$ 1.875$s^{-1}$\cite{antoun06}\\
\hline
$k_{3c2c}=$ 0.034$s^{-1}$\cite{antoun06} & $k_{3p2p}=$ 0.42$ s^{-1} $ \cite{antoun06}&$k_{3nc2nc}=$ 0.034$s^{-1}$\cite{antoun06}\\
\hline 
$k_{3'c2c}=$ 0.034$s^{-1}$\cite{antoun06} & $k_{3'p2p}=$  0.42$ s^{-1} $ \cite{antoun06}&$k_{3'nc2nc}=$ 0.034$ s^{-1} $\cite{antoun06}\\
\hline 
$k_{3c3'c}=$ 1.3$s^{-1}$\cite{wang15} & $k_{3p3'p}=$ 0.1$ s^{-1} $ &$k_{3nc3'nc}=$ 0.1$ s^{-1} $\\
\hline 
$k_{3'c3c}=$ 0.96$s^{-1}$\cite{wang15} & $k_{3'p3p}=$ 0.1$ s^{-1} $ &$k_{3'nc3nc}=$ 0.1 $s^{-1}$\\
\hline
$k_{3c4c}=$ 0.013$s^{-1}$ \cite{wang15} & $k_{3p4p}=$ 0.1$s^{-1}$ &$k_{3nc4nc}=$ 0.65$s^{-1}$ \cite{wang15}\\
\hline 
$k_{3'c4c}=$ 0.013$ s^{-1}$ \cite{wang15}  & $k_{3'p4p}=$ 0.1$s^{-1}$ & $k_{3'nc4nc}=$ 0.65$s^{-1}$ \cite{wang15}\\
\hline
$k_{4c3c}=$ 1.95$s^{-1}$\cite{wang15} &$k_{4p3p}=$ 0.015$s^{-1}$ &$k_{4nc3nc}=$ 0.63$s^{-1}$ \cite{wang15}\\
\hline
$k_{4c3'c}=$1.95$s^{-1}$ \cite{wang15} &$k_{4p3'p}=$ 0.015$s^{-1}$ &$k_{4nc3'nc}=$ 0.63$s^{-1}$ \cite{wang15}\\
\hline
$k_{3c5c}=$ 2.9$s^{-1}$ \cite{antounEMBO06} & $k_{3p5p}=$ 0.0071$s^{-1}$\cite{antoun06} &$k_{3nc5nc}=$ 0.06$s^{-1}$ \cite{wang15}\\
\hline 
$k_{3'c5c}=$ 2.9$ s^{-1}$ \cite{antounEMBO06}  & $k_{3'p5p}=$ 0.0071$s^{-1}$ \cite{antoun06}& $k_{3'nc5nc}=$ 0.06$s^{-1}$ \cite{wang15}\\
\hline
$k_{5c3c}=$ 1.12$s^{-1}$\cite{macdougall15} &$k_{5p3p}=$ 3$s^{-1}$ &$k_{5nc3nc}=$ 3$s^{-1}$\\
\hline
$k_{5c3'c}=$1.12$s^{-1}$ \cite{macdougall15}  &$k_{5p3'p}=$ 3$s^{-1}$ &$k_{5nc3'nc}=$ 3$s^{-1}$\\
\hline
$k_{5c6c}=$ 1.76$s^{-1}$ \cite{goyal15}& $k_{5p6p}=$1.76$s^{-1}$&$k_{5nc6nc}=$ 1.76$s^{-1}$\\
\hline
\end{tabular}
\end{center}
\end{table}

The rate constants that we use for the computer simulations of our model and for the graphical plots of the 
analytical results are given in Table \ref{table_parameter}; these have been collected from the published 
literature on experimental studies of translation initiation. All the available rate constants for association 
events are in $ \mu M^{-1}s^{-1} $ units, which are, effectively, second-order rate constants. 
Nonetheless, our theoretical model assumes that all transition rate constants are in the unit $ s^{-1} $ units. Thus, we have
converted all second-order rate constants into pseudo-first-order rate constants by calculating the product of the second-order rate constant by an assumed substrate concentration. For example, $ k_{12}=k_{12}[S] $, where [S] is the concentration of the substrate. In Table II we have listed the converted value of all the rate constants using $ [S]=0.15 \mu M $ \cite{milon12}.

\section{Results}

We began our analysis by first solving the master equations (see Section 1 of the Appendix), using the matrix method and determining the corresponding expressions for $ P_{i}(t) $. Detailed calculations and all of the 
analytical expressions are given in the supplementary material. 
For a given set of values of the model parameters (rate constants), results obtained from these exact analytical solutions are plotted, using continuous curves. For the same set of values of the parameters, the corresponding numerical 
data, obtained from the kinetic Monte Carlo simulations of the 
model have been plotted with discrete data points on the same graphs as the exact analytical solutions.
For the given set of values of the parameters, plotting the results in this manner allows a direct comparison between the results of the computer simulations of the model and the exact analytical solutions.

\begin{figure}[ht]
\centering
(a)\
\includegraphics[angle=-0,width=0.3\columnwidth]{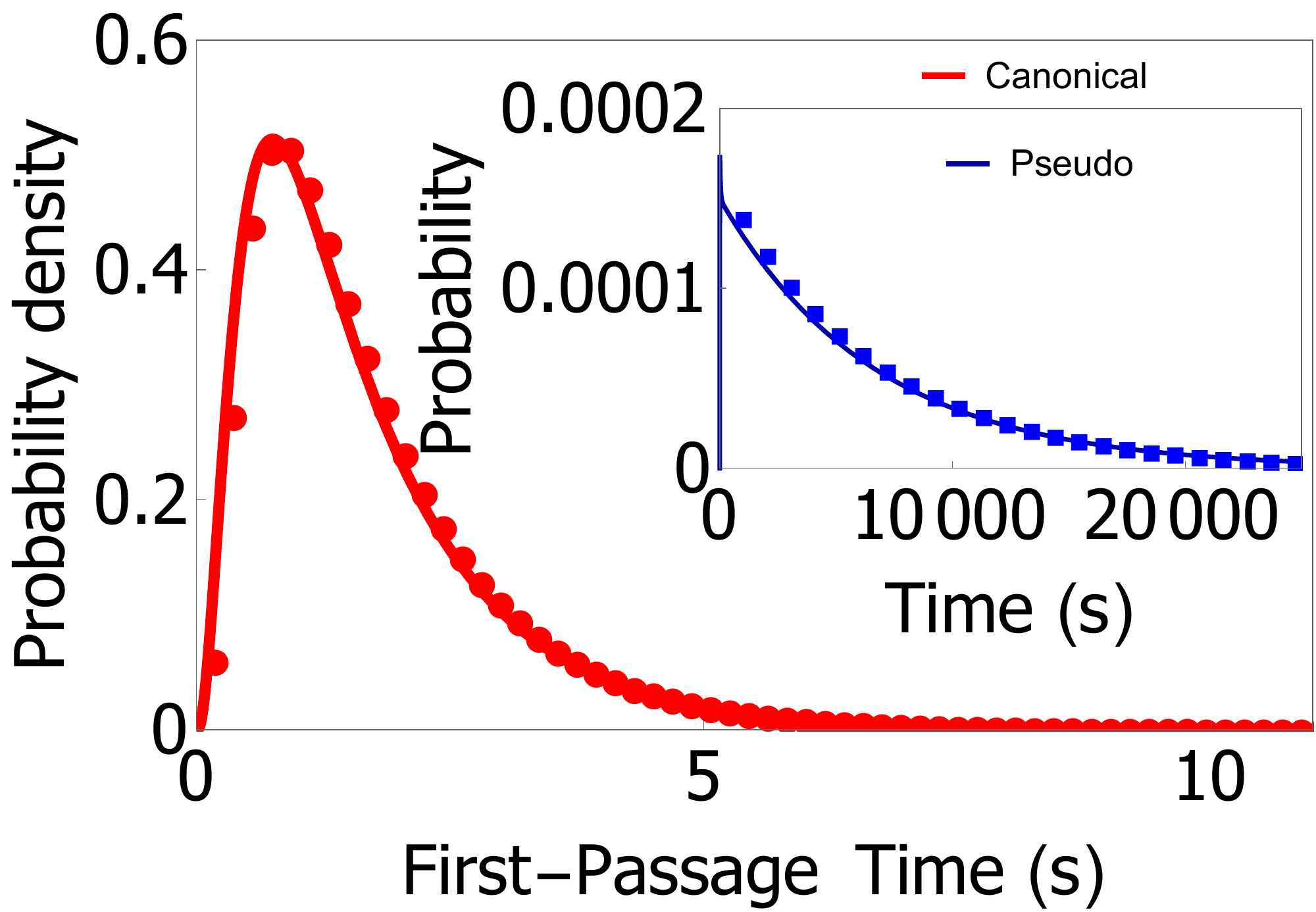}
(b)\
\includegraphics[angle=-0,width=0.3\columnwidth]{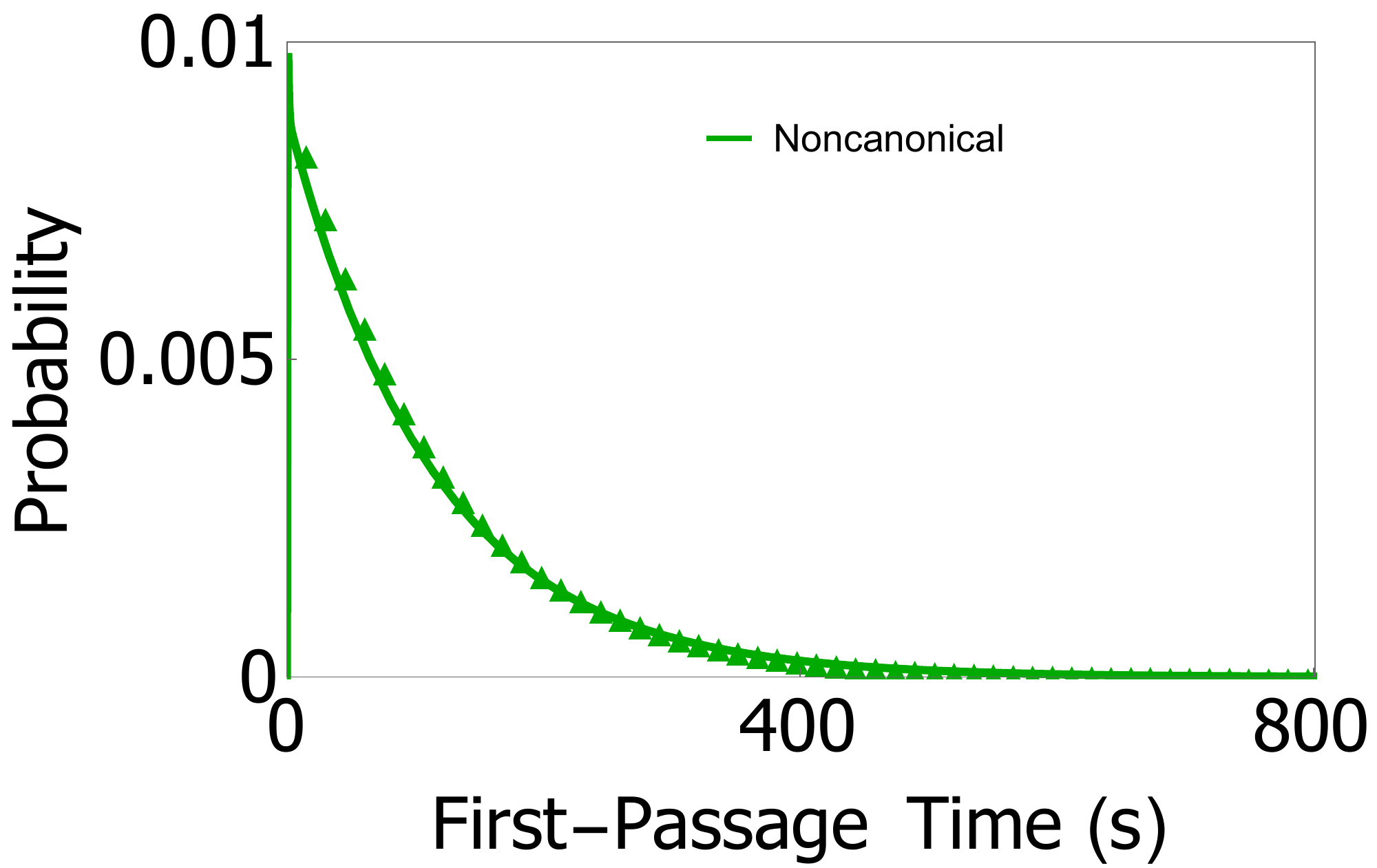}
(c)\
\includegraphics[angle=-0,width=0.3\columnwidth]{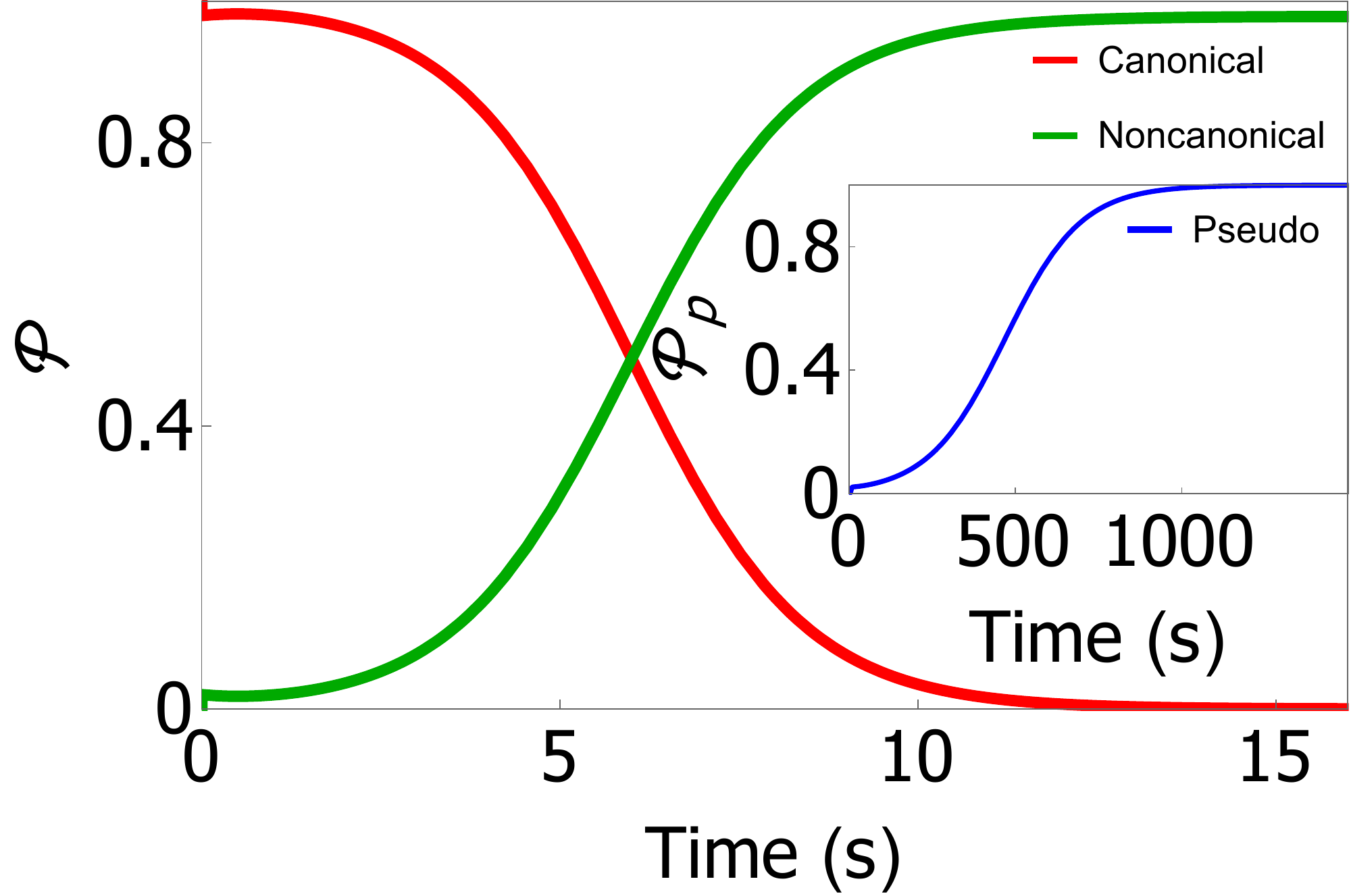}
\caption{(a) The First-passage time distribution for canonical 70S IC assembly. 
The first-passage time distribution for pseudo 70S IC assembly is shown in the inset.
 (b) The first-passage time distribution for non-canonical 70S IC assembly.
Discrete data points have been obtained from kinetic Monte Carlo simulations.
(c) The probabilities that the 70S ICs assembled at time $t$ is of the type canonical, pseudo or 
non-canonical are plotted as functions of time. 
Values of the parameters used for the calculation are listed in the Table \ref{table_used}.}
\label{fig_fptdist}
\end{figure}


The FPT distributions for assembly of the canonical and pseudo 70S ICs using the values of the parameters presented in Table II
are plotted in Figure \ref{fig_fptdist}(a) and (b). Because of the orders-of-magnitude longer times required for completion of 
pseudo 70S IC assembly, the pseudo 70S IC FPT distribution is shown separately in the inset of 
Figure \ref{fig_fptdist}(a). Clearly, for the set of parameter values used in these plots, the FPT distribution 
for canonical 70S IC assembly is sharper than that of non-canonical 70S IC assembly. In contrast, the FPT distribution for pseudo 70S 
IC assembly is much broader (on the order of 6 $ hr $) than those of canonical and non-canonical 70S 
IC assembly. 

In Table \ref{table_tavg}, the average time required for assembly of all three 70S ICs obtained from the exact analytical solutions of the model is compared with the data obtained form kinetic Monte Carlo simulations. 
As might be expected, assembly of a canonical 70S IC is orders-of-magnitude faster than formation of pseudo and non-canonical 70S ICs. Interestingly, the average time required for pseudo 70S IC assembly is very long
($ \sim 2 $ hr), compared to that of either canonical or non-canonical 70S IC assembly, suggesting that this is the rarest error that should be observed either \textit{in vitro} or \textit{in vivo}. This is presumably because all three IFs are involved in tRNA selection while only IF1 and IF3 are involved in discriminating against non-canonical start codons.
These observations for the average times of assembly are consistent with the most probable values in the graphical plots in Figure \ref{fig_fptdist}(a) and (b).

The time required to initiate  protein synthesis \textit{in vivo} and \textit{in vitro} have been reported as 4 s  \cite{ingraham83,farewell98} and 2 s \cite{antounEMBO06}, respectively values that are very similar to that which we have obtained here (1.65 s).
 The reported rate of 50S subunit docking during canonical 70S IC assembly (2.9 $ s^{-1} $\cite{antounEMBO06}) is much larger than
 that reported for pseudo 70S IC assembly (0.0071 $s^{-1}$\cite{antoun06}) and non-canonical 70S IC assembly (0.06 $ s^{-1} $\cite{wang15}),
consequently resulting in the differences we observe in the average times of assembly among all three 70S ICs.
In our model, the frequencies of selecting an incorrect tRNA or codon are given by the rate constants $ k_{p} $
and $ k_{nc} $. Interestingly, O'Connor and co-workers  \cite{connor01} have reported that the error frequency of selecting an incorrect codon (\textit{i.e.}, of forming a non-canonical 70S IC) is in the range of $ 10^{-3} $-$ 10^{-2} $, whereas we have calculated the
frequency of selecting an incorrect codon and forming a non-canonical 70S IC  assembly is 8.8 $ \times 10^{-3} $, a value that is in very good agreement with the range reported by O'Connor and co-workers. We find that the randomness parameter is much less than one for canonical 70S IC assembly ( $r_{c} < $ 1), whereas it is much closer to 1 for both pseudo and non-canonical 70S IC assembly ( $r_{p}$ and $r_{nc} $ , respectively). This observation suggests that there are fewer rate limiting steps in the canonical 70S IC assembly pathway than there are in the pseudo and non-canonical 70S IC assembly pathways.

The probabilities that, at any instant in time, a canonical, pseudo, or non-canonical 70S IC has been assembled, ${\cal P}_{c}(t)$, ${\cal P}_{p}(t)$ , and  ${\cal P}_{nc}(t)$, respectively, are plotted in Figure \ref{fig_fptdist}(c). These plots demonstrate that ${\cal P}_{c}(t)$  increases as a function of time, whereas both ${\cal P}_{p}(t)$ and ${\cal P}_{nc}(t)$ decrease as a function of time. 
 
\begin{table}[h]
\caption{Average time required for assembly of canonical, pseudo, and non-canonical 70S ICs \label{table_tavg}}
\begin{center}
\begin{tabular}{|c|c|c|c|}
\hline 
\multicolumn{1}{|p{2cm}}{\centering Parameter}
&\multicolumn{1}{|p{2cm}}{\centering Analytical}
&\multicolumn{1}{|p{2cm}|}{\centering Simulation}\\
\hline
$<t_c>$  & 1.65 $ s $ & 1.65 $ s $\\
\hline
$<t_{ps}>$ & 6658.54 $ s $ & 6654.9 $ s $ \\
\hline
$<t_{nc}>$ & 112.92 $ s $ & 112.86 $ s $\\
\hline
$r_c$  & 0.56  & 0.56 \\
\hline
$r_{ps}$ & 1.0  & 1.0\\
\hline
$r_{nc}$ & 0.99  & 0.99 \\
\hline
\end{tabular}

\end{center}
\end{table}
.

\section{Summary and conclusion}
Translation initiation is a multi-step, kinetically controlled process. In this article, we have 
developed  a theoretical model for the stochastic kinetics of translation initiation that captures
all of the major pathways, and the key steps along those pathways, for 70S IC assembly.
These stochastic kinetics have been formulated 
in terms of master equations. Solving these equations for the appropriate initial 
conditions, we have derived exact analytical results that characterize the 
assembly of canonical, pseudo, and non-canonical 70S ICs.
 We have also carried out direct kinetic Monte Carlo simulations of 
the model; the numerical data obtained from these simulations are in excellent agreement 
with the analytical results for the same set of values of the parameters. 
The sequence with which the IFs bind to the 30S subunit at the earlier stages of translation initiation and 
the conformational transitions of the IFs, tRNA, 30S subunit, and 50S subunit at the later stages of translation initiation
 are not described in detail in the current version of the model reported here.
Nonetheless, this model can be easily extended in the future by incorporating such further details of translation 
initiation at the cost of increasing the number of states and the inter-state transitions.

\section*{Acknowledgements}  
This work is supported by ``Prof. S. Sampath Chair'' professorship (D.C.),  a J.C. Bose National Fellowship (D.C.), National Institutes of Health (NIH) grant number GM 084288 (R.L.G), American Cancer Society Postdoctoral Fellowship number 125201 (K.C.), and NIH grant number R01 GM 29169 (J.F.).
\clearpage
\section*{Author contributions} 

D.C., R.L.G. and J.F. conceived the research; D.C. and D.G. derived the equations, performed the calculations and
simulations, and analyzed the data; D.G., K.C., and D.C. interpreted the results; D.G., K.C., J.F., R.L.G., and D.C.
wrote the manuscript; all five authors approved the final manuscript.

\begin{appendix}
\section{Master equations} 
\noindent {\it{Master equation for canonical, pseudo, and non-canonical 70S IC assembly}}:
\begin{eqnarray}
\frac{dP_{1}(t)}{dt}&=&-(k_{12c}+k_{12p}++k_{12nc})P_{1}(t)\nonumber\\
\frac{dP_{2m}(t)}{dt}&=&-(k_{2m3m}+k_{2m3'm})P_{2m}(t)+k_{12m}P_{1}(t)+k_{3m2m}P_{3m}(t)+k_{3'm2m}P_{3'm}(t)\nonumber\\
\frac{dP_{3m}(t)}{dt}&=&-(k_{3m2m}+k_{3m3'm}+k_{3m4m}+k_{3m5m})P_{3m}(t)+k_{2m3m}P_{2m}(t)+k_{3'm3m}P_{3'm}(t)\nonumber\\
&+&k_{4m3m}P_{4m}(t)+k_{5m3m}P_{5m}(t)\nonumber\\
\frac{dP_{3'm}(t)}{dt}&=&-(k_{3'm2m}+k_{3'm3m}+k_{3'm4m}+k_{3'm5m})P_{3'm}(t)+k_{2m3'm}P_{2m}(t)+k_{3m3'm}P_{3m}(t)\nonumber\\
 &+&k_{4m3'm}P_{4m}(t)+k_{5m3'm}P_{5m}(t)\nonumber\\
\frac{dP_{4m}(t)}{dt}&=&-(k_{4m3m}+k_{4m3'm})P_{4m}(t)+k_{3m4m}P_{3m}(t)+k_{3'm4m}P_{3'm}(t)\nonumber\\
\frac{dP_{5m}(t)}{dt}&=&-(k_{5m3m}+k_{5m3'm}+k_{5m6m})P_{5m}(t)+k_{3m5m}P_{3m}(t)+k_{3'm5m}P_{3'm}(t)\nonumber\\
\frac{dP_{6m}(t)}{dt}&=&k_{5m6m}P_{5m}(t)
\label{eq-master}
\end{eqnarray}
Now, by replacing \textit{m} by \textit{c, p,} or \textit{ nc} we can generate the master 
equations corresponding to canonical (\textit{m = c}), pseudo (\textit{m = p}), and non-canonical (\textit{m = nc}) 70S IC assembly

\section{First-passage times} 

The exact analytical expressions for the FPT distributions in the model depicted in Figure \ref{fig_model} are too long to be reproduced here; those are given in the Supplementary Information (SI). In the specific situation where the rate constants are assigned the numerical values given in Table \ref{table_used}, the FPT distributions are:
\begin{eqnarray}
f_{c}(t)&=&k_{5c6c}P_{5c}(t)\nonumber\\
        &=&4.86164 (-0.00164214 (1.17214 e^{-50.55 t}) + 1.07807\times 10^{-16} (0.00305401 e^{-7.81209 t}) \nonumber\\
        &+& 2.29629\times 10^{-17} (- 4.17097\times 10^{-6} e^{-7.76589 t}) + 0.87647 (0.412059 e^{-6.10705 t})\nonumber\\
        &+& 4.21646\times 10^{-15} (0.0768827 e^{-5.207 t}) + 0.866568 (- 0.00384965 e^{-3.90077 t}) \nonumber\\
        &-& 3.1498\times 10^{-16} (- 0.414465 e^{-3.79523 t}) - 0.84743 (0.726712 e^{-3.75461 t})\nonumber\\ 
        &-& 2.30305\times 10^{-16} (- 0.235286 e^{-1.89782 t}) + 5.92961\times 10^{-17} (- 7.53036\times 10^{-17} e^{-0.944 t})\nonumber\\
        &-& 0.790692 (- 0.32875 e^{-0.834563 t}) - 5.58734\times 10^{-15} (1.00213\times 10^{-15} e^{-0.7271 t})\nonumber\\
        &-& 1.60422\times 10^{-15} (0.0461257 e^{-0.580494 t}) -9.34218\times 10^{-17} (- 0.165401 e^{-0.0365675 t})\nonumber\\
         &+& 1.41576\times 10^{-16} (- 0.157824 e^{-0.0088676 t}) - 5.86036\times 10^{-17} (0.244105 e^{-0.00015009 t}))\nonumber\\
f_{p}(t)&=&k_{5p6p}P_{5p}(t)\nonumber\\
      &=&4.86164 (-6.77053\times 10^{-8} (1.17214 e^{-50.55 t}) + 1.36923\times 10^{-15} (0.00305401 e^{-7.81209 t})\nonumber\\
       &-&0.861924 (- 4.17097\times 10^{-6} e^{-7.76589 t}) -8.28588\times 10^{-17} (0.412059 e^{-6.10705 t})\nonumber\\
       &-&9.90501\times 10^{-17} (0.0768827 e^{-5.207 t}) + 4.87183\times 10^{-17} (- 0.00384965 e^{-3.90077 t})\nonumber\\
       &+&5.12745\times 10^{-18} (- 0.414465 e^{-3.79523 t}) -2.69575\times 10^{-17} ( 0.726712 e^{-3.75461 t})\nonumber\\
       &-&1.65771\times 10^{-17} (- 0.235286 e^{-1.89782 t}) - 1.45105\times 10^{-18} (- 7.53036\times 10^{-17} e^{-0.944 t})\nonumber\\
       &+&1.16821\times 10^{-16} (- 0.32875 e^{-0.834563 t})- 5.81008\times 10^{-17} (1.00213\times 10^{-15} e^{-0.7271 t})\nonumber\\
       &-&0.000902986 (0.0461257 e^{-0.580494 t}) - 0.0000440266 (- 0.165401 e^{-0.0365675 t}) + 1.74155\times 10^{-21}\nonumber\\
       && (0.157824 e^{-0.0088676 t}) + 0.000126393 (0.244105 e^{-0.00015009 t}))\nonumber\\
f_{nc}(t)&=&k_{5nc6nc}P_{5nc}(t)\nonumber\\
         &=& 6.37763 (-0.0000268627 (1.17214 e^{-50.55 t}) + 0.849945 (+ 0.00305401 e^{-7.81209 t}) - 1.11985\times 10^{-14}\nonumber\\
         &&(- 4.17097\times 10^{-6} e^{-7.76589 t}) -9.32006\times 10^{-17} (0.412059 e^{-6.10705 t}) - 1.39124\times 10^{-16}\nonumber\\
         &&(0.0768827 e^{-5.207 t}) - 1.39785\times 10^{-16} (- 0.00384965 e^{-3.90077 t}) + 0.0142295(- 0.414465 e^{-3.79523 t})\nonumber\\
         &-&2.17402\times 10^{-16} (0.726712 e^{-3.75461 t}) - 0.00825034 (- 0.235286 e^{-1.89782 t}) - 4.53702\times 10^{-18}\nonumber\\
         &&(- 7.53036\times 10^{-17} e^{-0.944 t}) - 6.04146\times 10^{-17} (- 0.32875 e^{-0.834563 t}) - 8.53461\times 10^{-18}\nonumber\\
         &&(1.00213\times 10^{-15} e^{-0.7271 t}) + 8.41874\times 10^{-17} (0.0461257 e^{-0.580494 t}) + 4.59984\times 10^{-17}\nonumber\\
         &&(- 0.165401 e^{-0.0365675 t}) - 0.00882116 (- 0.157824 e^{-0.0088676 t})+4.25025\times 10^{-16} (0.244105 e^{-0.00015009 t}))\nonumber \\
\label{eq_fptd}
\end{eqnarray}
where the subscripts \textit{c},\textit{nc} and \textit{ps} refer to the canonical, pseudo, and non-canonical 70S ICs, respectively. These FPT distributions are plotted in Figure \ref{fig_fptdist}(a).

\end{appendix}




\begin{thebibliography}{99}

\bibitem{frank11} Frank, J. Introduction. In Molecular Machines in Biology: Workshop of the Cell; Frank, J., Ed.; Cambridge University Press:
New York, 2011; pp 1-3.

\bibitem{nobel} Ramström, O. presentation speech for the 2016 Nobel Prize in Chemistry, Stockholm Concert Hall, 2016. 

\bibitem{tanford01} Tanford, C.; Reynolds, J. J.; nature's robots; Oxford University Press, 2001.

\bibitem{spirin00} Spirin A S, Ribosomes ; Berlin: Springer, 2000. 

\bibitem{rodnina12} Rodnina, V.; Wintermeyer, W.; Green, R. (Eds.) Ribosomes: Structure, Function, and Dynamics; Springer, 2012. 

\bibitem{misteli01} Misteli, T. The concept of self-organization in cellular architecture. J. Cell Biol. {\bf 2001}, 155, 181-185.

\bibitem{bressloff14} Bressloff, P. C. Stochastic Processes in Cell Biology; Springer, 2014.

\bibitem{redner01} Redner, S. A Guide to First-Passage Processes; Cambridge University Press: Cambridge, 2001.

\bibitem{metzler14} Metzler, R.; Oshanin, G.; Redner, S. First Passage Phenomena and their Applications; World Scientific, 2014. 

\bibitem{biswas16} Iyer-Biswas, S.; Zilman, A. First-Passage Processes in Cellular Biology. Adv. Chem. Phys. {\bf 2016}, 160, 261-306.




\bibitem{gualerzi15} Gualerzi, C. O.; Pon, C. L.; Cell. Mol. Life Sci. {\bf 2015}, 72, 4341-4367 (2015).

\bibitem{milon12} Milon, P.; Maracci, C.; Filonava, L.; Gualerzi, C. O.; Rodnina, M. V. Real-time assembly landscape of bacterial 30S translation initiation complex. Nat. Struct. Mol. Biol. {\bf 2012}, 19, 609-616.





\bibitem{antoun06} Antoun, A.; Pavlov, M. Y.; Lovmar, M.; Ehrenberg, M. How Initiation Factors Maximize the Accuracy of tRNA Selection in Initiation of Bacterial Protein Synthesis. Moll. Cell. {\bf 2006}, 23, 183-193.

\bibitem{antounEMBO06} Antoun, A.; Pavlov, M. Y.; Lovmar, M.; Ehrenberg, M. How initiation factors tune the rate of initiation of protein synthesis in bacteria. EMBO J. {\bf 2006}, 25, 2539-2550.

\bibitem{milon08} Milon, P.; Konevega, A. L.; Gualerzi, C. O.; Rodnina, M. V. Kinetic Checkpoint at a Late Step in Translation Initiation. Mol. Cell. {\bf 2008}, 30, 712-720.

\bibitem{simonetti08}  Simonetti, A.; Marzi, S.; Myasnikov, A. G.; Fabbretti, A.; Yusupov, M.; Gualerzi, C.; Klaholz, B.P. Structure of the 30S translation initiation complex. Nature. {\bf 2008},  455 , 416-420.

\bibitem{hussain16} Hussain, T.; Llácer, J. L.; Wimberly, B. T.; Kieft, J. S.; Ramakrishnan, V. Large-scale movements of IF3 and tRNA during bacterial translation initiation. Cell. {\bf 2016}, 167, 133-144.

\bibitem{zorzet10} Zorzet, A.; Pavlov, M. Y.; Nilsson, A. I.; Ehrenberg, M.; Andersson, D. I. Error-prone initiation factor 2 mutations reduce the fitness cost of antibiotic resistance. Mol. Microbiol. {\bf 2010}, 75, 1299-1313.

\bibitem{pavlov11} Pavlov, M. Y.; Zorzet, A.; Andersson, D. I.; Ehrenberg, M. Activation of initiation factor 2 by ligands and mutations for rapid docking of ribosomal subunits. EMBO J. {\bf 2011}, 30, 289-301.

\bibitem{caban17} Caban, K.; Pavlov, M. Y.; Ehrenberg, M.; Gonzalez, R. L. Jr. A conformational switch in initiation factor 2 controls the fidelity of translation initiation in bacteria. Nat. Commun. {\bf 2017}, 8, 1-11.

\bibitem{sussman96} Sussman, J. K.; Simons, E. L.; Simons, R. W. Escherichia coli translation initiation factor 3 discriminates the initiation codon in vivo. Mol. Microbiol. {\bf 1996}, 21, 347-360.

\bibitem{sacerdot96} Sacerdot, C.; Chiaruttini, C.; Engst, K.; Graffe, M.; Milet, M.; Mathy, N.; Dondon, J.; Springer, M. The role of the AUU initiation codon in the negative feedback regulation of the gene for translation initiation factor IF3 in Escherichia coli. Mol. Microbiol. {\bf 1996}, 21, 331-346.

\bibitem{haggerty97} Haggerty, T. J.; Lovett, S. T. IF3-mediated suppression of a GUA initiation codon mutation in the recJ gene of Escherichia coli. J. Bacteriol {\bf 1997}, 179, 6705-6713.

\bibitem{olsson96} Olsson, C. L.; Graffe, M.; Springer, M.; Hershey, J. W. B. Physiological effects of translation initiation factor IF3 and ribosomal protein L20 limitation inEscherichia coli. Mol. Gen. Genet. {\bf 1996}, 250, 705-714.

\bibitem{qin12} Qin, D.; Liu, Q.; Devaraj, A.; Fredrick, K. Role of helix 44 of 16S rRNA in the fidelity of translation initiation. RNA. {\bf 2012}, 18, 485-495.

\bibitem{grigoriadou07} Grigoriadou, C.; Marzi, S.; Kirillov, S.; Gualerzi, C. O.; Cooperman, B. S. A quantitative kinetic scheme for 70 S translation initiation complex formation. J. Mol. Biol. {\bf 2007}, 373, 562-572.

\bibitem{samhita13} Samhita, L.; Virumae, K.; Remme, J. Initiation with Elongator tRNAs. J. Bacteriol. {\bf 2013}, 195, 4202-4209.

\bibitem{guillon92} Guillon, J. M.; Mechulam, Y.; Schmitter, J. M.; Blanquet, S.; Fayat, G. Disruption of the gene for Met-tRNA(fMet) formyltransferase severely impairs growth of Escherichia coli. J. Bacteriol. {\bf 1992}, 174, 4294-4301.

\bibitem{basu07} Basu, A.; Chowdhury, D. Traffic of interacting ribosomes: Effects of single-machine mechanochemistry on protein synthesis. Phys. Rev. E {\bf 2007}, 75, 021902-11. 

\bibitem{garai09} Garai, A.; Chowdhury, D.; Chowdhury, D.; Ramakrishnan, T. V. Stochastic kinetics of ribosomes: single motor properties and collective behavior. Phys. Rev. E {\bf 2009} , 80, 011908-15.

\bibitem{sharma11} Sharma, A. K.; Chowdhury, D. Distribution of dwell times of a ribosome: effects of infidelity, kinetic proofreading and ribosome crowding. Phys. Biol. {\bf 2011}, 8, 26005-11.

\bibitem{dutta17} Dutta, A.; Chowdhury, D. A generalized Michaelis-Menten equation in protein synthesis: effects of mis-charged cognate tRNA and mis-reading of codon. Bull Math Biol. {\bf 2017}, 79, 1005-1027.


\bibitem{bar07} Bar, N. S.; Morris, D. R. Dynamic Model of the Process of Protein Synthesis in Eukaryotic Cells. Bull Math Biol. {\bf 2007}, 69, 361-393.

\bibitem{zouridis07} Zouridis, H.; Hatzimanikatis, V. A model for protein translation: polysome self-organization leads to maximum protein synthesis rates. Biophys. J. {\bf 2007}, 92,717-730.



\bibitem{wang15} Wang, J.; Caban, K.; Gonzalez, R. L. Jr. Ribosomal initiation complex-driven changes in the stability and dynamics of initiation factor 2 regulate the fidelity of translation initiation. J. Mol. Biol. {\bf 2015}, 427, 1819-1834.

\bibitem{studer06} Studer, S. M.; Joseph, S. Unfolding of mRNA Secondary Structure by the Bacterial Translation Initiation Complex. Mol. Cell. {\bf 2006}, 22, 105-115.

\bibitem{macdougall15} MacDougall, D. D.; Gonzalez, R. L. Jr. Translation initiation factor 3 regulates switching between
different modes of ribosomal subunit joining. J. Mol. Biol. {\bf 2015}, 427, 1801-1818.





\bibitem{goyal15} Goyal, A.; Belardinelli, R.; Maracci, C.; Milon, P.; Rodnina, M. V. Directional transition from initiation to elongation in bacterial translation. Nucleic Acids Res. {\bf 2015}, 43, 10700–10712.


\bibitem{ingraham83} Ingraham, J. L.; Maaloe, O.; Neidhardt, F. C.; Sunderland, MA 01375: Sinauer Associates Inc (1983).

\bibitem{farewell98} Farewell, A.; Neidhardt, F. C. Effect of Temperature on In Vivo Protein Synthetic Capacity in Escherichia coli. J. Bacteriol. {\bf 1998}, 180, 4704-4710.

\bibitem{connor01} O’Connor, M.; Gregory, S. T.; Rajbhandary, U. L.; Dahlberg, A. E. Altered discrimination of start codons and initiator tRNAs by mutant initiation factor 3. RNA, {\bf 2001}, 7, 969-978. 


\end{thebibliography}
\end{document}